\documentstyle[aps,pre,epsfig,floats]{revtex}

\begin{document}

\draft
\twocolumn[\hsize\textwidth\columnwidth\hsize\csname@twocolumnfalse%
\endcsname

\title{Ordering dynamics of the driven lattice gas model}
\author{E. Levine, Y. Kafri, and D. Mukamel\\[2mm]}

\address{Department of Physics of Complex Systems,
Weizmann Institute of Science, Rehovot 76100, Israel}
\date{\today}
\maketitle

\begin{abstract}
The evolution of a two-dimensional driven lattice-gas model is
studied on an $L_x \times L_y$ lattice. Scaling arguments and
extensive numerical simulations are used to show that starting
from random initial configuration the model evolves via two
stages: ({\em a}) an early stage in which alternating stripes of
particles and vacancies are formed along the direction $y$ of the
driving field, and  ({\em b}) a stripe coarsening stage, in which
the number of stripes is reduced and their average width
increases. The number of stripes formed at the end of the first
stage is shown to be a function of $L_x/L_y^\phi$, with $\phi
\simeq 0.2$. Thus, depending on this parameter, the resulting
state could be either single or multi striped. In the second,
stripe coarsening stage, the coarsening time is found to be
proportional to $L_y$, becoming infinitely long in the
thermodynamic limit. This implies that the multi striped state is
thermodynamically {\em stable}. The results put previous studies
of the model in a more general framework.
\end{abstract}

\pacs{PACS numbers: 05.40.-a, 05.70.Ln, 64.60.My, 64.60.Cn}]
\section{Introduction}
Driven diffusive systems have been extensively studied in recent
years. They serve as a fruitful framework for studying the
statistical mechanics of systems far from thermal equilibrium.
Driven by an external field these systems reach a steady state
with a non-vanishing current and as such do not satisfy detailed
balance. Studies of these models have revealed many differences
between such systems and systems in thermal equilibrium. For
example, several one-dimensional driven diffusive systems with
local dynamics exhibit long range order and spontaneous symmetry
breaking. Such phenomena can not occur in thermal equilibrium when
the interactions are short ranged \cite{David,Evans}.

Many studies of driven diffusive systems have focused on a driven
lattice gas (Ising) model. The model was introduced by Katz,
Lebowitz, and Spohn \cite{KLS} and is often referred to as the
``standard model''. In $d=2$ dimensions the model is defined on an
$L_x \times L_y$ lattice. Each of the lattice sites $i$, is either
occupied by a particle or is vacant. A macroscopic configuration
is characterized by a set of occupation numbers $\{ n_i \}$ where
$n_i=0,1$ represents a vacant or an occupied site, respectively.
Usually the model is studied with an equal number of occupied and
vacant sites. An energy ${\cal H}=-\sum_{\langle ij \rangle} n_i
n_j$ is associated with each configuration. Here the sum is over
$\langle ij \rangle$ nearest neighbor sites. The energy represents
an attractive interaction between the particles. An external drive
is introduced through a field $E$ which biases the motion of the
particles in the $-y$ direction. Imposing periodic boundary
conditions in this direction results in a current of particles
through the system along the field direction. Specifically the
dynamics of the model is defined through the exchange of nearest
neighbor particles with a rate
\begin{equation}
W = \min \left\{1,\exp( -\beta \Delta {\cal H} - E \Delta y)
\right\} \label{rate}.
\end{equation}
Here $\beta$ is an inverse temperature-like parameter, and $\Delta
y=(-1,0,1)$ for a particle attempting to hop along, orthogonal to,
or against the direction of the driving field. The energy
difference between the two configurations after and before the
particle exchange is denoted by $\Delta {\cal H}$.

The model has been studied extensively for nearly two decades
\cite{DL17}. Monte-Carlo simulations suggest that the $(T,E)$
phase diagram of the model is composed of two phases: A high
temperature disordered phase in which the particle density is
homogeneous, and a low temperature phase in which the system
orders and phase separates into high density and low density
regimes. It was found that in this phase the particles evolve
towards a striped structure parallel to the direction of the
driving field. Numerical studies indicate that a slow coarsening
takes place in this state  \cite{MV,YRHJ,ALLZ,RY}. As the
magnitude of the driving field is increased, the transition
temperature between the two phases increases and saturates at
about $1.41 \;T_O$ \cite{Leu}, where $T_O$ is the Onsager
temperature corresponding to $E=0$.

Recent Monte-Carlo simulations of this model suggest that the
evolution of the striped phase is rather complex. For a square
system the stripes are found not to coarsen in the thermodynamic
limit, yielding a multi-striped ordered state. This phase was
termed extraordinary or ``stringy'' \cite{ZSS}. On the other hand
systems with large aspect ration, $L_y \gg L_x$, were found to
evolve toward a single stripe phase.

\begin{figure}
\begin{center}
\epsfxsize 8.5  cm \epsfbox{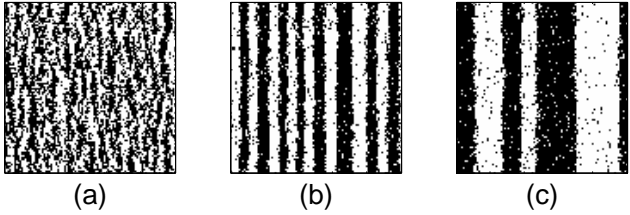}\vspace{0.2cm} \caption{A
typical evolution of a system of size $L_x=L_y=100$ from a random
initial condition. Configurations from times (a) 50, (b) 2000 and
(c) 500,000 Monte-Carlo sweeps are shown. Here $\beta=2$ and
$E=\infty$. One can clearly observe the two steps of the
coarsening process described in the text.} \label{config}
\end{center}
\end{figure}

In order to get a better understanding of the nature of the
ordered phase of the driven lattice gas model we carry out in this
paper a finite size scaling analysis of the evolution process
starting from a fully disordered state. We find that the model
evolves via
two stages:\\
\noindent ({\em a}) an early {\em stripe formation} stage in which
stripes are formed from the initially disordered state; and\\
\noindent ({\em b}) a {\em stripe coarsening} stage in which the
multi-stripe configuration formed in the early stage coarsen by
reducing the number of stripes and increasing their average width.
A typical evolution of such a system is shown in Fig.
\ref{config}.

Our studies yield two main results:\\
\noindent ($1$) The number of stripes which are formed at the end
of the initial stripe formation stage strongly depends on the
aspect ratio of the system. In particular we find that the number
of stripes $m$ scales as $m \sim L_x / L_y^{\phi}$, with $\phi
\simeq 0.2$. This implies that for narrow systems ($L_x /
L_y^{\phi} \lesssim 1$) a single stripe is formed at the end of
the first stage, while for wide systems ($L_x / L_y^{\phi} \gg 1$)
the resulting structure is multi-striped.\\
\noindent ($2$) Simple arguments are presented to show that during
the stripe coarsening stage the average width of the stripes grows
with time as $(t/L_y)^{1/3}$. This behavior is verified by
extensive numerical simulations. Therefore, the coarsening of the
stripes becomes slower as the system size in the direction of the
drive is increased. This implies that in the thermodynamic limit a
multiple striped configuration is in fact stable. We note that
similar phenomena of arrested striped configurations have been
observed in previous studies of coarsening of other models with
striped structures perpendicular to the direction of the drive
\cite{EKLM}.

The paper is organized as follows: In Section II the stripe
formation stage is discussed. Section III considers the stripe
coarsening stage. We end with a summary and discussion of the
implications of our results to other related works in Section IV.

\section{The stripe formation stage}

The evolution of the driven lattice gas model in the early stripe
formation stage has received some attention \cite{YRHJ,ALLZ,RY}.
Numerical simulations indicate that the domain growth process
which takes place in this stage is highly anisotropic. The typical
domain size in the direction of the drive and the direction
perpendicular to it grow differently. In particular it has been
observed \cite{YRHJ,ALLZ} that the typical domain size parallel to
the drive grows roughly as $\ell_y \sim t^{\varphi_\parallel}$
with $\varphi_\parallel \simeq 1$, while the typical domain size
perpendicular to the drive grows roughly as $\ell_x \sim
t^{\varphi_\perp},\;\;\varphi_\perp \simeq 0.2$. This behavior is
very different from that of a non-driven system evolving towards
equilibrium. It is well known that in such a system, when the
dynamics is conserving, as is the case here, the average linear
domain size $\xi$ grows as $t^{1/3}$ \cite{BRAY}. The difference
in behavior is due to the inherent anisotropy induced by the
drive.

The number of stripes formed in the system at the end of the
stripe formation stage can be estimated using the results
described above. For a stripe to form in the system the size of a
domain along the direction of the drive $\ell_y(t)$ must be of the
order of the system size $L_y$. Since $\ell_y(t) \sim
t^{\varphi_\parallel}$ the time for this to occur $t_s$ scales as
$t_s \sim L_y^{1/\varphi_\parallel}$. At this time the typical
domain size perpendicular to the drive is
\begin{equation}
\ell_x(t_s) \sim t_s^{\varphi_\perp} \sim L_y^{\varphi_\perp /
\varphi_\parallel}.
\end{equation}
Thus, the number of stripes formed, $m$, scales as
\begin{equation}
m \sim \frac{L_x}{\ell_x(t_s)} \sim \frac{L_x}{L_y^{\phi}},
\label{mscaling}
\end{equation}
where $\phi={\varphi_\perp / \varphi_\parallel}$. Using the
estimates for the exponents $\varphi_\perp$ and
$\varphi_\parallel$ one has $\phi \simeq 0.2$.

Specifically, for a square system, where $L_x = L_y \equiv L$,
Eq.~\ref{mscaling} implies $m \sim L^{1-\phi}$. Since $\phi < 1$
we find that the number of stripes grows as the system size is
increased, and one always reaches a multi-striped state.
The stripe density, $m/L_x$, vanishes in the thermodynamic limit.
In fact, this result is independent of the exact value of the
exponent $\phi$ as long as $\varphi_\parallel > \varphi_\perp$.

\begin{figure}
\begin{center}
\epsfysize 4.2cm \null \hspace{0.2cm} \epsfbox{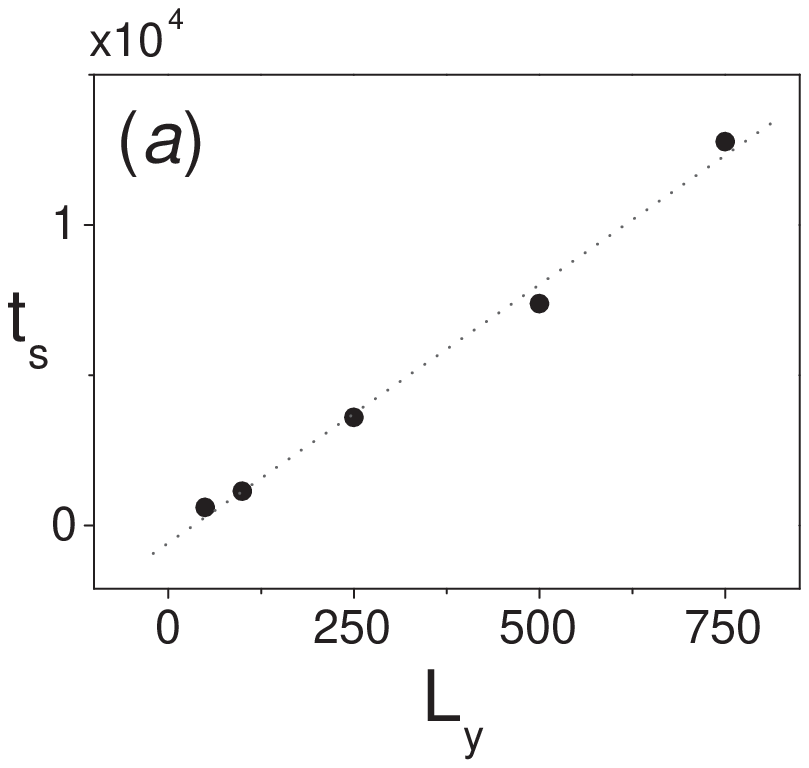} \\
\vspace{-0.7cm} \epsfysize 4 cm \epsfbox{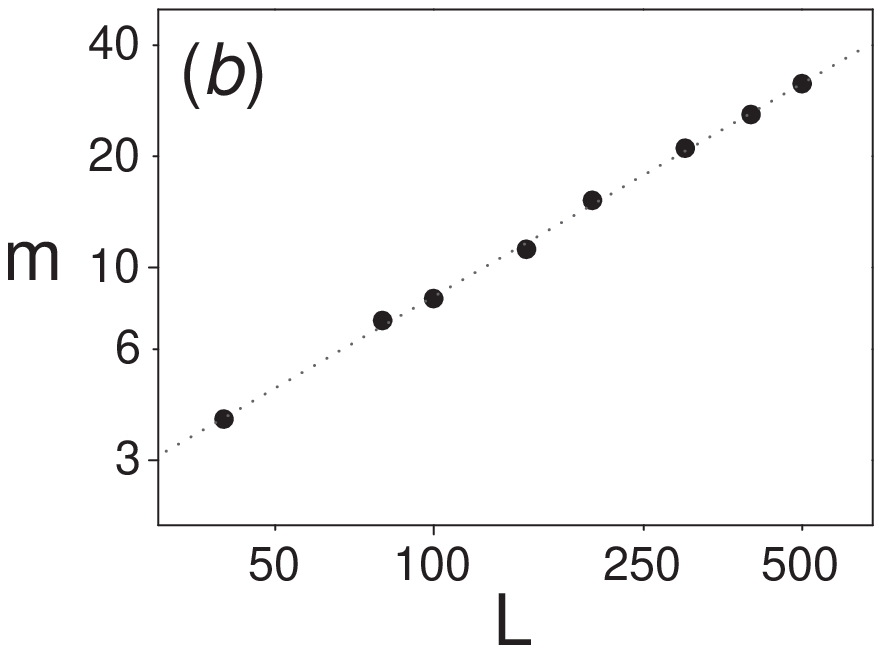} \caption{({\it
a}) The stripe formation time $t_s$ (in Monte-Carlo sweeps)
plotted against $L_y$, for systems with $L_x=100$. The behavior is
consistent with $\varphi_\parallel \simeq 1$. ({\it b}) The number
of stripes formed at the end of the first stage in square systems
of various sizes $L\equiv L_x = L_y$, plotted on a log-log scale.
The straight line corresponds to $m \sim L^{0.82}$. }
\label{mstripes}
\end{center}
\end{figure}

To verify these results Monte-Carlo simulations are performed for
various system sizes, starting from a random initial condition.
The Monte-Carlo procedure we use is standard: At each time step a
pair of neighboring sites is chosen randomly and updated according
to the rate $W$ given in (\ref{rate}). Throughout the paper we use
$E=\infty$ and $\beta=2$, for which the system is ordered. We have
checked that the main features of this study are unchanged for
other values of the parameters as long as the system is in the
ordered phase. We first verify the growth law of $t_s$ with $L_y$.
In order to evaluate $t_s$ the equal-time correlation of two sites
at a distance $L_y/2$ in the drive direction is measured and
averaged over the sample. The time $t_s$ is estimated by the time
at which the average measured correlation reaches the value of
$0.4$. The results are shown in Fig.~\ref{mstripes}($a$). One can
see that the behavior of $t_s$ with $L_y$ is consistent with
$\varphi_\parallel \simeq 1$.

The number of stripes initially formed in the system is estimated
by performing a Fourier transform of the density in the $x$
direction and locating its first peak at a non-zero wave length.
This procedure is repeated 40 times for each system size. For
simplicity we consider only square systems. In
Fig.~\ref{mstripes}($b$) we plot the location of the peak for
square systems as a function of $L$. The fitted exponent for
slightly over a decade of system sizes gives $\phi \approx 0.18
\pm 0.05$ which fits rather well with the values predicted by the
argument $\phi \approx 0.2$. The linear dependence of $m$ on $L_x$
for non square systems is also verified through simulations which
are not shown in this paper.

In general, the number of stripes also depends on the magnitude of
the driving field and the temperature. One can write, based on
Eq.~\ref{mscaling}, $ m =A L_x/L_y^{\phi}$, where the amplitude
$A(E,\beta)$ is introduced. We find that the amplitude is an
increasing function of both the magnitude of the driving field $E$
and the inverse temperature $\beta$. We note that when $L_y^\phi
\geq A L_x$ the number of stripes already at the end of the first
stage is expected to be one.

\section{The stripe coarsening stage}
We now turn to the second stage of the ordering process, namely
the coarsening of the stripes formed in the early stage. We
present a simple argument suggesting that the time $t$ in which
stripes coarsen scales linearly with the system size parallel to
the drive, $L_y$. Moreover, we show that the average stripe width
$\ell(t)$ scales with time as $(t/L_y)^{1/3}$. The fact that the
characteristic time associated with the coarsening process scales
linearly with $L_y$ implies that in the thermodynamic limit, where
$L_y \to \infty$, the coarsening time becomes infinite and thus
the multi-stripe structure exists as a thermodynamically stable
state. Our argument relies on two main features of the driven
system: ($i$) the fact that the ordered domains, namely the
stripes, are of the size of the system, and ($ii$) the smoothness
of the domain walls bounding the stripes.  This last feature has
been shown to be a result of the drive
\cite{LMV,Leung,YMHAJ,LeuZia}. In contrast to the non-driven
two-dimensional Ising model, where the domain walls are rough,
here the driving field 
makes the domain walls smooth.

We proceed by considering a striped state composed of alternating
stripes of particles and vacancies with average width $\ell$.
Neighboring stripes of particles interact with each other by an
exchange of particles. Since the boundaries of the stripes are
smooth, the lateral distance that particles have to travel in
order to move from one stripe to the other is of the order of
$\ell$. To estimate the coarsening time we assume that within a
stripe of vacancies the density of particles is low enough so that
the particles may be considered as non-interacting. This
assumption is qualitatively supported by the configurations
observed in simulations (see, e.g, Fig. \ref{config}). When a
particle reaches the boundary it is absorbed in the neighboring
particle stripe. Thus the lateral motion of the particles within a
stripe of vacancies can be considered as a one-dimensional random
walk in the $x$ direction with two absorbing walls located at
$x=0$ and $x=\ell$. This problem is known as the gambler's ruin
problem \cite{FELLER}. The probability of such a particle to move
from $0$ to $\ell$ is given by $p(\ell) \sim 1/\ell$.

For the width of a stripe to decrease by one lattice spacing it
has to lose $L_y$ particles. Due to the right-left symmetry of the
problem, the particle currents from one stripe to the other are
balanced on average. Therefore a net transfer of particles from
one stripe to another is only due to fluctuations in the lateral
current. The net excess in the number of particles transferred at
a time interval $t$ is then proportional to $\sqrt{L_y p(\ell)
t}$. For one stripe to shrink and disappear $\ell L_y$ particles
must be transferred so that
\begin{equation}
\sqrt{L_y p(\ell) t} \sim \ell L_y \;.
\end{equation}
Combining this result with $p(\ell) \sim 1/\ell$ one finds that
the average stripe width in the system grows as
\begin{equation}
\ell (t) \sim \left( \frac{t}{L_y} \right)^{1/3} . \label{scaling}
\end{equation}
This suggests that the coarsening time scales with $L_y$, yielding
a stable striped structure in the thermodynamic limit.

The scaling form (\ref{scaling}) may be verified numerically by
studying the two point particle-particle correlation function. To
carry out this analysis we note that in an isotropic system
without a driving field, the coarsening process is characterized
by a single length scale $\xi(t)$, which could be the linear size
of the growing domains. In this case the two point
particle-particle correlation function obeys a scaling form
\cite{BRAY}
\begin{equation}
C(r,t)=g \left( \frac{r}{\xi(t)} \right) , \label{correlation}
\end{equation}
where $r$ is the distance between two points. Driven systems, on
the other hand, are non-isotropic, and correlations along the
drive and perpendicular to it behave differently. The typical
length scale perpendicular to the drive is given by
(\ref{scaling}). Thus we expect the correlation function in the
$x$ direction to be of the form
\begin{equation}
C_\perp(x,t)=g_\perp \left( \frac{x}{\left(t/L_y\right)^{1/3}}
\right) . \label{mainresult}
\end{equation}
The asymptotic behavior of $g_\perp(z)$ for $z \to 0$ is expected
to obey Porod's law, which states that $g_\perp(z) = 1/2 - \eta z$
with some constant $\eta$. For $z \to \infty$ one should have
$g_\perp(z) \to 1/4$.

We now turn to describe numerical studies which support our
results. Note that the scaling variable in (\ref{mainresult})
involves three parameters. All three parameters are varied in our
numerical studies. This is a demanding computational task, and a
good collapse of the data is a strong conformation of the scaling
analysis. In these studies an initial striped configuration along
the drive direction $y$ is considered, and its evolution is
simulated. The two point correlation function $C_\perp(x,t)$ is
then calculated, and is shown to obey the scaling form
(\ref{mainresult}). The widths of the stripes in the initial
configuration are randomly chosen from a Poisson distribution with
a mean width $\ell_0$. The simulations are performed for lattices
of three different sizes: $960 \times 8$, $800 \times 16$ and $960
\times 32$. We consider several values of $L_x$ to demonstrate
that this parameter does not play an important role in the
process. The mean width of the stripes in the initial
configurations is taken to be $\ell_0=4$. The two-point
particle-particle correlation function $C_\perp(x,t)$ is measured
and averaged over 110, 75 and 54 simulations for $L_y=8$, $16$ and
$32$, respectively.

\begin{figure}
\begin{center}
\epsfxsize 8  cm \epsfbox{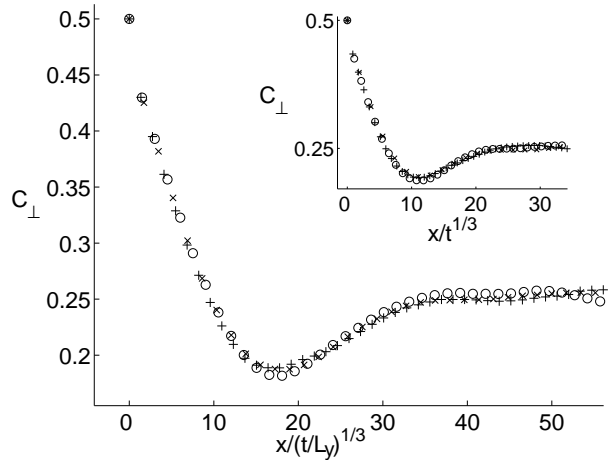} \vspace{0.2cm}
\caption{The two-point particle-particle correlation function
$C_\perp$ is plotted as a function of the scaling variable
$x/(t/L_y)^{1/3}$, for systems of size $960 \times 8$ (marked by
$\times$), $800 \times 16$ ($\circ$), $960 \times 32$ ($+$). The
times of measurement are chosen arbitrarily ($t=1, 3 , 8 \times
10^6$ Monte-Carlo sweeps, respectively). In the inset, $C_\perp$
as a function of $x/t^{1/3}$ is shown for a system of size $960
\times 8$ and times $t=0.2$($\times$), $1$($\circ$), $2$($+$)
$\times 10^6$ Monte-Carlo sweeps. } \label{collapse}
\end{center}
\end{figure}

The scaling form (\ref{mainresult}) suggests that data collapse
should take place with respect to the three variables $x,t$ and
$L_y$. This collapse is checked in two steps. First we consider
$L_y =8$ and show that $C_\perp$ is a function of $x/t^{1/3}$ as
expected. This is demonstrated in the inset of Fig.
\ref{collapse}. Similar results are obtained for the other system
sizes as well.

Next, we verify the full scaling form (\ref{mainresult}). In Fig.
\ref{collapse} correlation functions for the three different
system sizes are plotted. For each system size the correlation
function is evaluated for arbitrarily chosen $t$ and the data is
then plotted as a function of the scaling variable
$x/(t/L_y)^{1/3}$. Again, the quality of the data collapse
supports our main result. Although computation time limits us to a
relatively small systems, we believe the quality of the data backs
our scaling argument.

\section{Discussion}
The evolution of the driven lattice gas model was considered
starting from a random initial configuration. We have shown, using
simple scaling arguments and extensive numerical simulations, that
the evolution proceeds via two stages: an early, stripe formation
stage in which stripes of the size of the system are formed,
followed by a second stage in which the stripes coarsen. While the
first stage lasts $t_s \sim L_y$, the system evolves towards a
single stripe configuration in the second stage at a time of order
$\sim L_x^3 L_y$. This is a result of the fact that the typical
width of stripes in the coarsening stage scales with time as
$\ell(t) \sim (t/L_y)^{1/3}$. This result indicates that the
coarsening time of multi-striped configurations scales with the
system length $L_y$, suggesting that these configurations exist as
stable states in the thermodynamic limit $L_y \rightarrow \infty$.

Thus, starting from a random initial configuration, the system
evolves to one of two types of states, depending on its aspect
ratio. For $L_x / L_y^{\phi} \leq 1$ ($\phi \simeq 0.2$) the
stripe formation stage leads directly to a single stripe state,
while for $L_x / L_y^{\phi} \gg 1$ multi-striped states are
reached. The coarsening process of these states proceeds with a
time scale proportional to $L_y$.

These results put in a more general framework previous studies of
this model which considered either the early stages
\cite{YRHJ,ALLZ,RY} of the evolution or the nature of the steady
state \cite{DL17}. A recent study of a square system has shown
\cite{ZSS} that a multi striped state (termed "stringy") is
reached from a random initial condition. It was suggested that
this state is stable. Our studies indicate that this is indeed the
case for an infinitely large system. However we expect a finite
system to coarsen to a single striped state at a time of the order
of $L_x^3 L_y$. The fact that the steady state of a system with a
small aspect ration, $L_x/L_y$, was found to be composed of a
single stripe is consistent with our scaling picture.

Finally, note that the slow coarsening of the stripes is a direct
consequence of stripes spanning the entire system. This is a
result of the existence of the drive, and is expected to be valid
also in higher dimensions. It would be interesting to study such
processes in high-dimensional systems. We note, however, that
already in two dimensions the computational effort was
considerable.

Acknowledgments: We thank Mustansir Barma for helpful discussions.
The support of the Israeli Science Foundation is gratefully
acknowledged.


\end{document}